\documentclass{ws-ijmpa-}
\def\mb{\medbreak}
\newcommand{\be}{\begin{equation}}
\newcommand{\ee}{\end{equation}}
\def\lbar{{\mathchar'26\mskip-9mu\lambda_e}} 
\def \lsa {\rlap {\lower 3.5 pt \hbox {$\mathchar \sim$}} \raise 1pt \hbox {$<$}}
\def \rsa {\rlap {\lower 3.5 pt \hbox {$\mathchar \sim$}} \raise 1pt \hbox {$>$}}
\begin{document}
\markboth{A.W.Weidemann, P.Chen, C.-K.Ng}
{Plasma Lens Backgrounds at a Future Linear Collider}

\catchline{}{}{}

 \begin{flushright}
  SLAC-PUB-9207\\
  April 2002
 \end{flushright}

\vskip 2cm 
\begin{center}
 {\Large{\bf Plasma Lens Backgrounds at a Future Linear Collider
\footnote{Work supported by the U.S. Department of Energy 
under Contracts DE--FG02--95ER40910 (SC) and DE-AC03-76SF00515 (SLAC).}
}}
\end{center}
\vskip 1cm 
\begin{center}
{\bf Achim W. Weidemann}\\%
 Department of Physics,\\ University of South Carolina,\\
Columbia, SC 29208 \\
\vskip 0.6cm
{\bf  Pisin Chen, Cho-Kuen Ng}\\%
Stanford Linear Accelerator Center\\
 Stanford, CA 94309
\end{center}
\vfill
\begin{abstract}
\noindent
A 'plasma lens' might be used to enhance the luminosity of future 
linear colliders. However, its utility for this purpose 
depends largely on the potential backgrounds that may be induced
by the insertion of such a device in the interaction region
of the detector.In this note we identify different sources of such backgrounds,
calculate their event rates from the elementary interaction processes,
and evaluate their effects on the major parts of
a hypothetical Next Linear Collider (NLC) detector.   
For plasma lens parameters which give a factor of seven enhancement of the 
luminosity, and using the NLC design for beam parameters as a reference, 
we find that the background yields are fairly high, 
and require further study and improvements in detector technology to
avoid their impact.
\end{abstract}
\vskip 2cm
\begin{center}
{\em Contributed to the 4th International Workshop,\\ 
 Electron-Electron Interactions at TeV Energies
\\ Santa Cruz, California,  December 7 - 9, 2001}
\end{center}
\vskip 3cm
\title{PLASMA LENS BACKGROUNDS AT A FUTURE LINEAR COLLIDER
\footnote{Work supported by the U.S. Department of Energy 
under Contracts DE--FG02--95ER40910 (SC) and DE-AC03-76SF00515 (SLAC).}
}
\author{\footnotesize ACHIM W. WEIDEMANN\footnote{
Mail Address: SLAC, MS 94, P.O.B. 20450, Stanford, CA 94309.}}
\address{Department of Physics, University of South Carolina,\\
Columbia, SC 29208 
}

\author{P. CHEN and C.-K. NG }
\address{Stanford Linear Accelerator Center\\
 Stanford, CA 94309
}

\maketitle


\begin{abstract}
A 'plasma lens' might be used to enhance the luminosity of future 
linear colliders. However, its utility for this purpose 
depends largely on the potential backgrounds that may be induced
by the insertion of such a device in the interaction region
of the detector.In this note we identify different sources of such backgrounds,
calculate their event rates from the elementary interaction processes,
and evaluate their effects on the major parts of
a hypothetical Next Linear Collider (NLC) detector.   
For plasma lens parameters which give a factor of seven enhancement of the 
luminosity, and using the NLC design for beam parameters as a reference, 
we find that the background yields are fairly high, 
and require further study and improvements in detector technology to
avoid their impact.

\end{abstract}

\section{Introduction}
The use of a plasma lens as a final-focusing device to increase the
luminosity in $e^+ e^-$ 
linear colliders was proposed some time ago~\cite{chen1}.
Conventional quadrupole magnets for final on-line focusing
in high-energy accelerators have limited
focusing strengths (a few hundred MG/cm), while plasma lenses are able to
produce focusing strengths a few orders of magnitude higher, depending on
the plasma density.
A plasma-lens, final-focus system might be even more important to
achieve a reasonable luminosity in a proposed $e^- e^-$
collider.

Although the self-focusing effect of a plasma lens has been demonstrated
experimentally, most recently in experiments at the 
Final Focus Test Beam at SLAC~\cite{e150},
the implementation of a plasma lens near a particle detector in 
a collider requires
that the presence of the plasma will not introduce serious backgrounds
in the detector as a consequence of beam-plasma interactions.
In this note we identify different sources of such backgrounds,
calculate their event rates from the elementary interaction processes,
and evaluate their effects on the major parts of
a hypothetical Next Linear Collider (NLC) detector.

For plasma lens parameters which may give a factor seven enhancement of the 
luminosity, and using the NLC design for beam parameters as a reference, 
we find that the background yields are are fairly high, 
and require further study and improvements in detector technology 
or beam time structure to avoid their impact.

\section{Choice of Plasma Lens Parameters}

The background yield depends on both the beam and plasma lens parameters.
For the former, we take the NLC beam energy of 
$E_{beam} =  250$~GeV, and $N = 0.75 \times 10^{10}$ particles per bunch,
with $n_b = 190$ bunches  in a bunch train.\cite{NLC2001}

For the flat beams of NLC, with r.m.s.   beam sizes
$\sigma_{x}=245$nm, $\sigma_{y}=2.7$nm, $\sigma_{z}=110\mu$m,
\cite{NLC2001}
it will be difficult to focus further in $y$; 
however, in the $x$-direction there is some possibility of improvement, 
as will be shown here.

As an example, we consider then a plasma lens 
%
with a density $ n_{_P} = 2 \cdot 10^{18}$cm$^{-3},$
beginning at $ s_{0} = 1 cm$  from the final focus in absence of the lens.  

 Such a lens would have a focusing strength
 $ K = 2\pi r_{e} n_{_P} / \gamma \approx 7.24$ cm$^{-2}.$
At the entrance to the lens, the NLC beta functions are
changed from their value at the focus in absence of a lens,
$\beta^{*}_{x0}=  8$mm, $\beta^{*}_{y0}= 100\mu$m,
to

\begin{equation}
\beta_{x0} = \beta^{*}_{x0} \left(1+\left({ s_{0}\over\beta^{*}_{x0}}\right)^{2}\
 \right)
            = 2.05\mbox{cm}, 
\end{equation}
and similarly, $  \beta_{y0} = 100.01\mbox{cm}. $

The differential equation for the $\beta$-function in the plasma
lens as a function of the distance $s$ along the beam line,
$\beta ''' +4K\beta' + K'\beta = 0, $
can be solved with the initial conditions
(matching $\beta, \beta'$ at 
$s=s_{0},$ 
the entrance to the plasma lens),
yielding
\begin{equation}
\beta = {\beta_{0} \over 2} + {1 \over {2K\beta_{0}^{*}}}
      +\left({\beta_{0}\over2} - {1\over{2K\beta_{0}^{*}}}\right)
                             \cos[\nu(s-s_{0})]
      - { {2s_{0}} \over \nu\beta_{0}^{*}}\sin[\nu(s-s_{0})]
\end{equation}
where $\nu^{2} = 4K.$
We chose the thickness of the lens to be 
\begin{equation}
    l =  \pi / 2\nu  \approx 0.3 cm.
\end{equation}

The maximal reduction in $\beta^{*}$ can be found to be\cite{chen2}
\begin{equation}
  {\beta^{*} \over \beta_{0}^{*}} =
  { 1 \over { 1 + K\beta_{0}^{*}(\beta_{0} - \beta_{1}) }}
\end{equation}
where   $\beta_{0},\beta_{1}$ are the $\beta$-function at the entrance
and exit of the plasma lens, respectively.
With the parameters given, one finds
$ {\beta_{x}^{*} /     \beta_{x0}^{*}} \approx 1/9, $
and
$ {\beta_{y}^{*} /     \beta_{y0}^{*}} \approx 1/6.8. $
The factor by which the luminosity increases, 
given by the square roots of these ratios of $\beta$-functions, 
thus can be  as high as  7.8, but may be somewhat less in practice 
due to astigmatism.

Note that the plasma lens parameters chosen above
serve only as an illustration and are not necessarily optimized;
however, they motivate the choice of
the product of plasma density and plasma length,
$n_{_P}\cdot l = 0.6\cdot 10^{18}$cm$^{-2}$
as conservative for a study of backgrounds  produced 
by a plasma lens at the NLC.  
 
Here we consider only a plasma lens focusing
one beam just before the interaction point;
having a plasma lens for both beams would, of course, double the backgrounds.
One could also imagine having a suitably shaped plasma at the
interaction point itself.

\section{Sources of Backgrounds}
Background particles produced by a plasma lens are of three types,
namely, electrons/positrons, hadrons, and photons.
These particles originate from
different elementary physical processes underlying the interactions of the
incoming electron or positron beam with the plasma of the lens located
near the interaction point.
In this section, we outline all the processes responsible for the various
sources of backgrounds~\cite{baltay}.
Their cross sections and angular distributions are calculated in the next
section. From these results and the parameters shown,
the number of background particles can be determined
and their effect on the detector evaluated.

\subsection{Electrons}
Electrons and positrons arise from the scattering of the $e^+$ or $e^-$
beam with the electrons of the plasma. The processes for producing
electrons and positrons are then
\begin{description}
\item[\hspace{1em} ] Bhabha scattering, $e^+e^- \rightarrow e^+e^-, $
\item[\hspace{1em} ]
M{\o}ller scattering, $e^-e^- \rightarrow e^-e^-, $
\item[\hspace{1em} ]
elastic scattering, $e\,p \rightarrow e\,p, $
\item[\hspace{1em} ]
and inelastic scattering, $e\,p \rightarrow e\,X. $
\end{description}

\subsection{Hadrons}
The hadronic backgrounds come from the elastic and inelastic scattering of
the $e^+$ or $e^-$ beam, and also from
the inelastic scattering of photons (from
synchrotron radiation and bremsstrahlung),
with the protons in the plasma.
Hadrons are produced by
 
\begin{description}
\item[\hspace{1em} ] elastic scattering, $e\,p \rightarrow e\,p, $
\item[\hspace{1em} ]
and inelastic scattering, $e\,p \rightarrow e\,X, $ 
$\gamma \,p \rightarrow X. $
\end{description}

\subsection{Photons}	 
Photon backgrounds from a plasma are produced by
\begin{description}
\item[\hspace{1em} ]
Compton scattering, $ e\gamma \rightarrow e\gamma. $
\end{description}
\mb
\noindent
Four mechanisms  produce the incident photons:
synchrotron radiation at the final-focus
quadrupole, beamstrahlung from the colliding beams at the interaction point,
synchrotron radiation as a consequence of plasma focusing, and
bremsstrahlung of the $e^+$ or $e^-$ beam in the plasma. 
These photons are then scattered by plasma electrons into large angles. 

Note that all detectors at $e^+ e^-$ colliders have to cope
with photon backgrounds generated by synchrotron radiation 
and beamstrahlung scattering off masks or similar
structures, which is not considered here.

\section{Cross Sections and Event Rates}
 
The cross section of an electron or a photon scattered by the plasma for a
particular process is calculated by integrating its angular distribution
 
\be
 \sigma = \int_{\theta_c}^{\pi-\theta_c} {d\sigma \over d\Omega}
\, d\Omega,
\label{esigma}
\ee
 
\noindent
where $\theta_c$ is the angular cut for the scattered particles into the
detector, which is taken to be 150 mrad in our calculations. The number of
scattered particles $N_s$ for a bunch train is then given by
 
\be
N_s = n_b {\cal L} \sigma,
\label{ens}
\ee
 
\noindent
where $n_b$ is the number of bunches in a bunch train, and ${\cal L}$ is
the beam-plasma luminosity per bunch crossing, respectively.
The beam-plasma luminosity is given by
 
\be
{\cal L} = N n_{_P} l,
\label{elumi}
\ee
 
\noindent
where $N$ is the number of particles in a bunch, $n_{_P}$ is the plasma
density and $l$ is the thickness of the plasma. 

With the beam and plasma lens parameters of section 2 above,
the beam-plasma luminosity ${\cal L} $  is then
$ 4.5 \cdot 10^{27}$ cm$^{-2}$.
For an anticipated NLC luminosity\cite{NLC2001} of  
${\cal L}_{NLC} = 2 \cdot 10^{34}$ cm$^{-2}$s$^{-1}$,
the beam-beam luminosity per bunch crossing, 
${\cal L}_{NLC} / ( 120 Hz  \cdot n_b )$ is $8.8 \cdot 10^{29}$ cm$^{-2}$,
two orders of magnitude larger than the beam-plasma luminosity.
For a bunch train of $n_b = 190$
bunches, $n_b {\cal L} = 8.55 \cdot 10^{29}$ cm$^{-2}$.
 
\mb
While the energy of the electrons in the beam is monochromatic,
the energy of the photons incident on the plasma lens is not.
The photons originating from the final-focus quadrupole and the plasma
focusing, as well as those from the beamstrahlung at the interaction point,
follow the synchrotron-radiation spectrum,
while those from  bremsstrahlung in the plasma follow
the   Weizs\"{a}cker-Williams spectrum.
Thus the photon distribution can be written as
 
\be
n_{\gamma}(y) 
       = n_{_{SR}}^p(y) + n_{_{SR}}^q(y) + n_{_{Brem}}(y)+ n_{_{Beamst}}(y) ,
\label{ny}
\ee
 
\noindent
where $y = E_\gamma / E_{beam}$ is the ratio of the photon energy to the
beam energy, $n_{\gamma}(y)$ is the total photon spectrum, and
$n_{_{SR}}^p(y)$, $n_{_{SR}}^q(y)$,
$n_{_{Brem}}(y)$ and $n_{_{Beamst}}(y)$
are the contributions from plasma-lens focusing,
quadrupole focusing, bremsstrahlung and beamstrahlung, respectively.
The synchrotron radiation spectrum from a focusing system
can be approximated by the following expression~\cite{chen}:
 
\begin{equation}
 n_{_{SR}}^i(y) = {1 \over \pi} \Gamma({2\over3})
({\alpha d_{i}  \over \sqrt{3} \gamma \lbar})
(3 \Upsilon_i)^{2/3} y^{-2/3},
\quad\quad 0 \; \lsa \; y \; \lsa \; \Upsilon_i, \quad
i=p,q, b.
\label{esr}
\end{equation}
 
\noindent
Here $\lbar$ is the electron-Compton wavelength,
$\gamma$ is $E_{Beam} / m_{e}c^{2},$
$d_{i}$ is the length of the focusing element
($d_{q} = 1$ m for a quadrupole and $d_{p} =l = 3$~mm for a plasma lens),
and $\Upsilon_i =2 E_{c}^{i} /3 E_{beam}$,
where  $E_{c}^{i}$ is the synchrotron-radiation critical energy.
For  $ y > \Upsilon_{i}$,
the synchrotron-radiation power is exponentially small and
is  neglected here.
For a final-focus quadrupole,
$\Upsilon_q \sim 1 \cdot 10^{-5}$ and $E_c^{q} \sim 2.5$~MeV;
for the plasma lens considered here,
$\Upsilon_p \sim 0.05 $ and $E_c^{p} \sim 17.3$~GeV.
The beamstrahlung spectrum, $n_{_{Beamst}}(y),$  
can also be described by (9), that is
$n_{_{Beamst}}(y) = n_{_{SR}}^b(y) $, with parameters\cite{NLC2001}
$\Upsilon_p \sim 0.11, E_c^{p} \sim 41.25$~GeV and 
$ d_{q} = 2 \sqrt{3}\sigma_{z} =0.38 $mm.
This formula will overestimate the beamstrahlung spectrum, as the integral
over $n_{_{Beamst}}(y)$ with the upper limit 0.2 yields $\sim$ 3, rather
than the 1.2 photons per incident electron indicated in the 
2001 Report on the NLC\cite{NLC2001}.

\mb
The bremsstrahlung spectrum from beam electrons scattered in the
plasma is given by the
Weizs\"{a}cker-Williams spectrum~\cite{jackson}
 
\be
 n_{_{Brem}}(y) = {2 \alpha \over \pi y}
[ ln {2.246} + ln{ m_p \over m_e }  - {1 \over 2}  - ln{y} ],
\label{ebrem}
\ee
 
\noindent
where $m_p$ and $m_e$  are the proton and electron masses, respectively.
Then the angular distribution of the cross section for each Compton
process, after taking the photon spectrum into account, is given by
\be
{d\sigma \over d\Omega} = \int_0^{y_0}
{d\sigma({e\gamma \rightarrow e\gamma}) \over d\Omega}
n(y) dy,
\label{edsda}
\ee
 
\noindent
where $n(y)$ is either the synchrotron radiation or the bremsstrahlung
spectrum,
and $y_0$ is the value above which radiation can be neglected, which is
$\sim 1.5 \cdot 10^{-5}$, 0.15 and 1 for quadrupole focusing, plasma
focusing, and bremsstrahlung in the plasma, respectively, and 0.2 for
beamstrahlung.

The integrated cross sections and backgrounds from the different processes
are summarized in Table 1.
We now discuss in more detail the production of each type of background
particle.

\begin{table}[h]
\begin{center}
{{\bf Table 1:} Summary of background sources from a plasma lens in NLC for
a single beam crossing.
The cross sections $\sigma_{tot}$ are integrated as in Eq.~(11) and (5); 
energy cuts (of $ 4-100 keV $ , $ > 100 keV $) were
imposed in the calculation of particle numbers in the
last two columns; see Section 5.
}
\vskip 3pt
\begin{tabular}{|l|c|c|c|}
\hline
 &   $\sigma_{tot}$ (cm$^{-2})$
 &  Vertex & Drift
\\
Background source & $\vert\cos\theta\vert \leq 0.99$
 & detector  & chamber
\\ \hline
Bhabha and M{\o}ller $e^+, e^-$ & 0 & 0 & 0
\\
Elastic $ep$: e  & $0.103 \times 10^{-45}$ & negligible & negligible
\\
\hskip 0.8 truein
              p  & $0.613 \times 10^{-39}$ & negligible & negligible
\\
Inelastic $ep$: e & $0.132 \times 10^{-33}$ & negligible & negligible
\\
\hskip 0.915 truein
charged hadrons   & $0.396 \times 10^{-29}$ & 0.021  & 0.021
\\
Inelastic $\gamma p$: charged hadrons &  $0.372 \times 10^{-28}$ 
& 0.139 & 0.139
\\
Compton $\gamma$'s from quadrupole & $0.18 \times 10^{-24}$
& 270  & 380
\\
Compton $\gamma$'s from plasma focusing & $0.23 \times 10^{-24}$
& 290   & 580
\\
Compton $\gamma$'s from bremsstrahlung & $0.19 \times 10^{-23}$
&   970  &  480
\\
Compton $\gamma$'s from beamstrahlung & $0.52 \times 10^{-25}$
&   70  &  130
\\
\hline
\end{tabular}
\end{center}
 
\noindent

\end{table}

\subsection{Electrons} 
The number of particles scattered into the detector depends on the
angular acceptance, which is taken to be  150~mrad  to $\pi-150$~mrad
from the incident-beam direction.
When a 250-GeV  electron or positron
hits an electron at rest, the scattered particles
go in the forward direction within a cone of very small opening
angle about the direction of the incoming electron. Thus we
expect a very small number of electrons scattered into the detector when a
beam passes through the plasma lens.
Electron backgrounds are essentially zero for Bhabha and M{\o}ller
scatterings, since the scattered particles
come out within the cone specified by our angular cut.
 
The angular distributions of the cross sections of the electrons
and protons produced by elastic $ep$ scattering are shown in Fig.~1.
The cross
section for electron production  peaks highly in the forward direction.
The total cross section for the imposed angular cut
is $0.103 \cdot 10^{-45}$ cm$^{-2}$, and hence, the
contribution to the electron sources is negligibly small.

The angular distribution of  the scattered electrons for inelastic $ep$
scattering can be obtained by integrating
the following differential distribution~\cite{martin} over energy:
 
\be
{d\sigma \over dE'd\Omega} =
{\alpha^2 \over 4 E^2 \mbox{sin}^4{\theta \over 2}}
\{ W_2(\nu,q^2)\mbox{cos}^2{\theta \over 2} +
   2W_1(\nu,q^2)\mbox{sin}^2{\theta \over 2} \}.
\label{eepx}
\ee
 
\noindent
Here $E$ and $E'$ are the incoming and outgoing electron energies
respectively,
$\nu = E -E'$, $q^2$ is the momentum transfer squared,
and $\alpha$ is the fine-structure constant. The parametrizations
of the structure functions $W_1(\nu,q^2)$ and $W_2(\nu,q^2)$ are taken
from Ref.~\cite{kmrs}.
The angular distribution
of the cross section for inelastic $ep$ scattering is shown in Fig.~2(a).
Again, we
 see that the cross section  peaks highly in the forward direction.
The total cross section is $0.132 \cdot 10^{-33}$ cm$^{2}$,
which gives
$0.11 \cdot 10^{-3}$ electrons for a bunch train.
Hence it can be concluded
that the electron backgrounds due to the
presence of a plasma lens can be neglected.

\subsection{Hadrons}
While the electrons essentially move in the forward direction, the protons
from $ep$ elastic scattering are scattered into larger angles.
In Fig.~1 we see that the differential cross section is a few orders of
magnitude bigger than that for the electrons. Nevertheless, the
integrated cross section is
$0.613 \cdot 10^{-39}$ cm$^{2}$ and hence the contribution
to the proton sources is negligibly small.
\begin{figure}
\centerline{\psfig{file=aw_plbg_fig1.epsi, width=8cm}}
\vspace*{8pt}
\caption{Angular distributions of the cross sections in the laboratory frame 
for $ep$ elastic scattering. The energy of the incoming electron is 250 GeV
and the proton is at rest.
}
\end{figure}
For inelastic $ep$ scattering, the initial-state proton can
disintegrate into other particles; to properly take this effect into
account
we use HERWIG~\cite{herwig} to simulate the inelastic $ep$ reactions
of 250-GeV electrons with protons at rest.
The differential distribution of charged hadrons as a function of scattering
angle is shown in Fig.~2(b); it again peaks at  small angles.
The total cross section for the angular cut
of 150~mrad is found to be $0.396 \cdot 10^{-29}$ cm$^{2}$, which
corresponds to about 3.4 charged hadrons for a bunch train at NLC energy.
Thus, we expect the hadronic backgrounds from $ep$ scattering will not
pose problems to any component of a NLC detector.
 
For inelastic $\gamma p$
scattering, where the photon originates from one of the mechanisms
mentioned in a previous section,
we found from HERWIG simulations that the cross section does not
vary sensitively with the incoming photon energy. Hence, we considered
a monochromatic photon of 10~GeV interacting with a proton at rest
for estimating the cross section.
The integrated cross section is found to be
$0.372 \cdot 10^{-28}$ cm$^{2}$,
which is about 8 times larger than that of the $ep$ inelastic scattering.
This corresponds to 32 charged hadrons for a bunch train, and hence this
background is still very small.
 
\begin{figure}
\centerline{\psfig{file=aw_plbg_fig2.epsi, width=8cm}}
\vspace*{8pt}
\caption{
(a) Angular distribution of the cross section in the laboratory frame for
the scattered electrons in $ep$ inelastic scattering.
(b) HERWIG angular distribution of charged hadrons for 20,000 events for
$ep$ inelastic scattering. The
total cross section is normalized to $1.02 \cdot 10^{7}$ pb.
}
\end{figure}

\subsection{Photons} 
The photon backgrounds for the Compton process from the four photon
sources depend on the photon spectrum.
When the photon energy is much greater than the electron rest mass,
the Compton cross section has a peak in the forward direction.
When the photon energy is comparable-to or less-than the electron mass,
the scattering is reduced to Thomson scattering
and the angular dependence of the cross section
varies as a dipole distribution. The angular distributions of the cross
sections of the Compton process from the four photon sources
are shown in Fig.~3. 
\begin{figure}
\centerline{\psfig{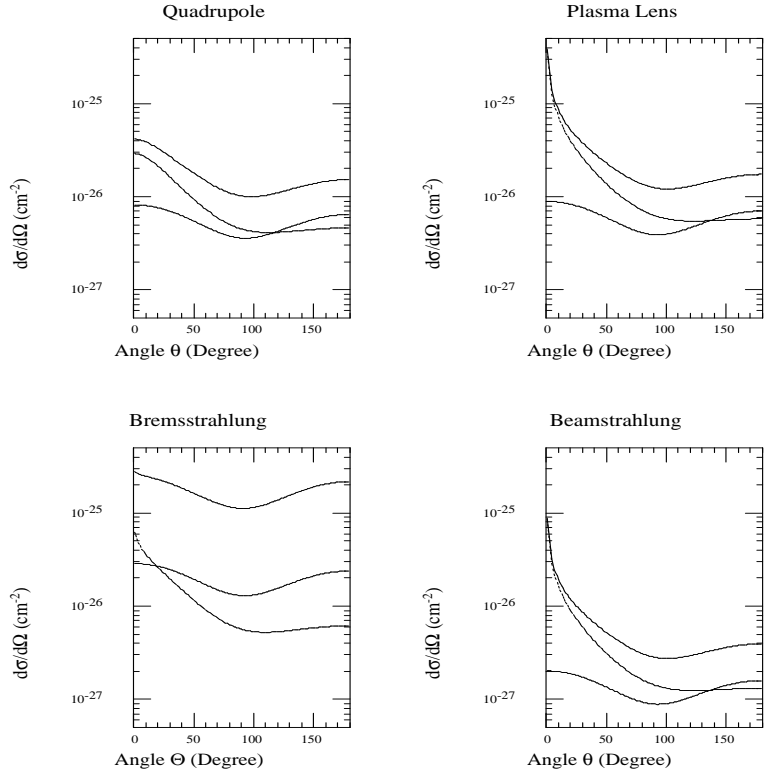}}
\vspace*{8pt}
\caption{
Angular distributions of the Compton cross sections from the four sources
of photons: quadrupole focusing, plasma-lens focusing.
bremsstrahlung in plasma,beamstrahlung.
The  topmost
solid lines are integrated over the whole energy range, the next lower ones
for the energy range between 4 keV and 100 keV 
(relevant for the vertex detector),
and the lines lowest (at right-hand side of plot) are
for photons of energy greater than 100 keV (relevant for the
drift chamber).
}
\end{figure}
The bremsstrahlung
spectrum is very soft; the angular distribution of this process is quite
flat: it looks like a dipole distribution.
The typical energy of the photon for synchrotron radiation
from final quadrupole focusing is
still small (on the order of one MeV) and hence, the
distribution does not vary drastically with angle although it peaks
moderately in the forward direction. Plasma focusing is much stronger
than  quadrupole focusing and the photon spectrum is expected to be
harder with the typical energy on the order of 10 GeV.
The cross section  peaks in the forward direction and then drops rapidly
at large angles.
The total number of photons of all energies
for a bunch train is $\sim$ 1,990,000, or about 10,500 for a single beam
crossing.(Photons above 1keV are $\sim$ 722,000 per bunch train,
or 3,800 per single beam crossing, still a considerable number.)
The photon backgrounds from $\gamma p$ inelastic
scattering are in general a few orders of magnitude smaller than those
from Compton scattering, and hence can be neglected.

\section{Detector Backgrounds}
As discussed in Section 4, the leptons and hadrons scattered into
the fiducial volume of a detector can be neglected.
The photons created by the different mechanisms
calculated above cover a wide range of energy.
Different components of future linear collider  detectors
are sensitive to certain
energy windows of the energy spectrum of the outcoming photons.
We now consider the photon backgrounds in
a vertex detector, a tracking device such as
a drift chamber or TPC, and a calorimeter of
an idealized NLC detector.
\subsection{Vertex detector} 
A description of the general characteristics of a pixel-based
vertex detector for the NLC can be found in Ref.~\cite{vertex}.
 
A minimum-ionizing charged particle will deposit 7.75 keV
and generate about 2,100 electron-hole pairs
in the, say, 20-$\mu$m-thick active Silicon of a vertex detector.
Photons above
50 keV, for example, are unlikely to convert in the very thin
structure of a barrel-vertex detector, which is expected to have a
thickness of as little as 0.11$\%$  of a radiation length\cite{vertex}.
Thus, to be conservative,
background photons of energies between, say, 4~keV and 100~keV
need to be considered here.
With these energy cuts, the number of background photons
for the vertex detector is $\sim$303,000 for a bunch train,
or 1600 per single-beam crossing,
compared with about $1.99\cdot 10^{6}$ photons per bunch train at all energies.
 
The angular distributions of
the cross sections from different mechanisms
are shown in Fig.~3; 
The cross section from bremsstrahlung photons is reduced by more than an
order of magnitude because of the cut at the lower end of the soft
Weizs\"{a}cker-Williams spectrum.
The cross sections from quadrupole and plasma focusing photons are
roughly reduced by a factor of 3.
 
For a vertex-detector barrel with a radius of 2.5~cm  and with
a length of 33 cm covering angles greater than 150 mrad
($\vert\cos\theta\vert \leq 0.99$),
the density of photon tracks per unit surface area is $\sim$ 5.8/mm$^2$
for a bunch train,
which is more than  the tolerable occupancy of $\sim$1/mm$^2$.
In comparison, the 'Snowmass 2001' detector design\cite{NLCPHYS2001}
forsees an occupancy of  $\sim$1/mm$^2$ per bunch train at a radius 
of 1.2 cm.
   
Note that the choice of $\vert\cos\theta\vert \leq 0.99$ for the vertex
detector is very generous, however
the number of background photons could only be slightly reduced if the 
angular coverage is limited to, for example, 
$\vert\cos\theta\vert \leq 0.90$. (Then  there are still $\sim$264,000
photons per bunch train, $\sim$1,400 per bunch crossing).
A better method would be to change the beam time structure, that
a vertex detector does not integrate noise over 190 pulses;
and to increase the readout speed of the device.
The noise integrated only over, say 30 bunches, would lead to an average 
occupancy of  $\sim$0.9/mm$^2$.

\subsection{Tracking}
The next layer in the detector may well be a large drift chamber or TPC
with about 10,000 sense wires.
Such a drift chamber can  easily tolerate 100 random background hits
(1\% occupancy).
Assuming a 1\%  conversion probability,
this  allows for 10,000 incident photons.
   Photons of very small energies, say $\lsa 10$ keV,
will be absorbed by the beam pipe and can be ignored.
  The conversion products of photons of less than, say 100 keV,
will not form track segments.
In a typical magnetic field they will form
tight loops  and deposit their energy locally.
Proper design of the readout
should account not only for the charge distribution typical of
minimum-ionizing particles, but also for the sometimes large local
depositions of energy from converting photons.
Thus we need only to consider photons of  energy greater than
100 keV scattered into a drift chamber or other tracking system.
 
The angular distributions of the
cross sections above 100 keV are also shown in Fig.~3.
They all peak in the forward direction as a result of the hard-photon
spectra from all the mechanisms.
The total number of background photons
with energy above 100~keV and scattering angle greater than 150~mrad
is about 299,000 per bunch train, or $\sim$1,600 per beam crossing.
Notice that we integrated the background over a whole bunch train.
Again, increasing the angular cut to 451 mr or
$\vert\cos\theta\vert \leq 0.90$ leads only to modest improvement
($\sim$240,000 photons per bunch train, $\sim$1,300 per bunch crossing).
  By the time NLC is built, we expect that progress in tracking technology
(e.g. 'bunch taggging'\cite{NLCPHYS2001})
will allow a better separation of tracks from adjacent bunch crossings,
such that only the background from a few adjacent bunch crossings
(instead of 190) needs to be considered.

\subsection{Calorimeter} 
As mentioned above, leptons and hadrons will not be scattered into the
main part of a detector in sufficient numbers to cause a problem for the
calorimeter of a hypothetical NLC detector.
Photons of very low energies will have been absorbed by
the beampipe, drift-chamber walls, and similar structures
before reaching the calorimeter.   A typical calorimeter
can  possibly see energies as low as a few MeV;
however, for any actual analysis,  a cut is made removing
clusters in a calorimeter with a total energy below, say,
100 MeV.
Our calculation shows that of the total flux per bunch train
of about  $1.99 \cdot 10^{6}$
photons of all energies in the angular region with
$\vert\cos\theta\vert\leq 0.99$,
5,950 have an energy above 100~MeV,(i.e. 31 per beam crossing,) 
and about 109,000 per bunch train, or $\sim$ 570 per single-beam 
crossing, have an energy between 1 and 100~MeV.
While some of the latter might be seen in the innermost
layer of electromagnetic calorimetry, they do not present
a large problem and are easily removed by a cluster cut, if isolated.
The energy added by them to a shower created by a high-energy
particle or jet is negligible compared to the intrinsic resolution
of a typical calorimeter.
Removal of the $\sim$31 higher-energy clusters per beam crossing 
requires more thought.
Due to the higher energy of the relevant photons, a larger angular cut 
helps more and removes about 40\% of these photons; in the region
$\vert\cos\theta\vert \leq 0.90$, about 3,500 photons per bunch train 
or 18 per bunch crossing remain with energy above 100 MeV.

Calorimeter backgrounds might also be attributed to individual bunch
crossings by the use of scintillation counters or similar fast timing
devices, so that again integration of the backgrounds over
a bunch train is excessively conservative.

\section{Summary}
We summarize our background calculations in Table 1, which shows
the various cross sections and particle fluxes for a single
bunch crossing from the different background sources.
using NLC parameters as a reference.
We have estimated only the sources of backgrounds from a plasma lens
for a generic particle detector,
but did not consider here at all the sources of backgrounds
generated by collisions without a plasma lens
that are common to any NLC detector,
such as those due to beam scrapings or synchrotron radiation
scattering off masks or similar structures, or the direct 
beamstrahlung from the opposing beam. (The compton scattering of the
beamstrahlung on the plasma lens was included.)

We have been rather conservative in our assumptions of 
the plasma lens parameters; presumably one with half the density 
(and thus producing half as much backgrounds), if properly optimized,
might still increase the luminosity by a factor of close to five.

Moreover, we have further assumed
that the resolution time for the detector is much longer than
1.4~ns, the separation between successive bunches in a bunch train,
but less than the $\sim$~8.3~ms 
separation between bunch trains.
and have integrated in our discussion the backgrounds 
over a train of 190 bunches. 
for a total time span of $\sim$270 ns.
  The thus-integrated backgrounds, overlayed over those produced
by collisions without a plasma lens, present indeed a challenge,
and require further study by a {\it bona fide} simulation
of both beams, plasma lens, and detectors.
Improved, especially faster, detector technology,
may be available by the time a future high-energy linear collider is
built; this, together with an optimized design (perhaps allowing for
lesser plasma density and a different time structure of the beam)
will possibly make the additional backgrounds created by a plasma
lens more acceptable, and as amenable as those anyway created by
high-energy collisions. 
Thus, the use of a plasma lens 
should certainly be considered carefully for the case that the
conventional focusing should not give the desired results, or
to increase the luminosity for certain physics channels
which might be easier separated from the backgrounds.


\section*{Acknowledgements}
 
\hskip 0.25 truein
We thank Clem Heusch for organizing again a stimulating workshop, and
T.~Barklow, C.J.S.~Damerell,  S.~Hertzbach,
T.~Markiewicz, S.~Rajagopalan, D.~Su,
and E.~Vella for useful discussions.


\end{document}